\documentclass[twocolumn, aps, prb, floatfix,superscriptaddress,nofootinbib, sortedaddress,longbibliography, notitlepage]{revtex4-2}

\usepackage[T1]{fontenc}
\usepackage{bm}
\usepackage[normalem]{ulem}
\usepackage{graphicx, color, soul}

\usepackage[usenames,dvipsnames, pdftex]{xcolor}

\usepackage{dcolumn}
\usepackage{amsmath,amsfonts, amssymb, amsthm, dsfont}

\usepackage{textcomp}
\usepackage{appendix}

\usepackage{bm}
\usepackage{tabularx, booktabs}
\usepackage{multirow}

\usepackage[colorlinks, breaklinks, 
            linkcolor=Blue,
            citecolor=Blue,
            urlcolor=Blue]{hyperref}           

\usepackage[all]{hypcap} 

\usepackage{enumitem}

\newcommand{\ra}{\rangle}
\newcommand{\la}{\langle}

\newcommand{\os}{\mathcal{O}^s_{q{=}\pi}}
\newcommand{\od}{\mathcal{O}^d_{q{=}\pi}}

\newcommand{\osO}{\mathcal{O}^s_{q{=}0}}
\newcommand{\odO}{\mathcal{O}^d_{q{=}0}}
\newcommand{\mD}{\mathcal{D}}

\begin{document}

\title{Fractonic Constraints and Magnetic Order in a Dipole-Conserving Spin Chain}
\author{Prabhakar}
\affiliation{Department of Physics, Indian Institute of Technology Bombay, Mumbai 400076, India}
\author{Giuseppe De Tomasi}
\affiliation{CeFEMA--LaPMET, Departamento de F\'{\i}sica, Instituto Superior T\'ecnico, Universidade de Lisboa, Portugal}

\author{Soumya Bera}
\affiliation{Department of Physics, Indian Institute of Technology Bombay, Mumbai 400076, India}

\date{\today}

\begin{abstract}
This work investigates the competition between dipole conservation, which imposes strong dynamical constraints and prevents the propagation of isolated spin excitations, and Ising-type interactions that favor ordering. Specifically, we explore the ground state phase diagram of a one-dimensional spin chain in the presence of both fractonic constraints and interactions.  Despite the kinetic constraints, the system stabilizes an antiferromagnetic dipole-ordered ground state, where the ordering occurs at the level of spin pairs rather than individual spins. At a large Ising interaction strength, the model undergoes a phase transition from a dipole-ordered phase to a spin antiferromagnetic phase. In contrast, for ferromagnetic Ising interactions, the model exhibits both antiferromagnetic and ferromagnetic dipole ordered phases. At sufficiently large negative interaction strength, the dipole ordered phase transitions to a ferromagnetic phase with conventional spin ferromagnetic order. To characterize these distinct phases, we employ density matrix renormalization group (DMRG) simulations alongside large-scale diagonalization. We analyze appropriate order parameters, along with features of the entanglement spectrum and dynamical spectral functions. In limiting cases, the observed transitions can be understood by mapping the dipole conserving model onto effective XXZ models in a restricted Hilbert space of composite spins.
\end{abstract}

\maketitle

\section{Introduction}
Symmetry is fundamental to understanding phases of matter. It determines which phases can exist and governs transitions between them. The spontaneous breaking of symmetry signals the emergence of a new order, giving rise to fundamentally distinct states with unique physical properties. In this context, multipolar-conserving systems-where not only charge but also higher moments such as the dipole are conserved-provide a particularly striking and fundamentally distinct setting in which enhanced symmetry constraints profoundly reshape the structure of quantum phases and their dynamics~\cite{hsf_1, khemani,morningstar,Moudgalya_RP22,SeidelPRL05,rahul_prr20, rahul_pre21}.

Dipole-conserving Hamiltonians define a class of constrained models that preserve the total dipole moment in addition to the total charge.
The additional conservation enforces correlated motion of particle pairs and restricts single-particle hopping, thereby providing a paradigmatic example of a fractonic system. These constraints give rise to a wide range of phenomena, including ergodicity-breaking Hilbert space fragmentation~\cite{hsf_1, khemani, hsf_2, hsf3, hsf4}, anomalous diffusion~\cite{tomasi_diff_prl, morningstar, bakr_prx, knap_prx25, rahul_prx19, rahul_prr20, rahul_pre21, piotr_prr21, Gliozzi_2023, Gliozzi_2026}, and several emergent phases~\cite{SeidelPRL05, BH_prb1, lake_prl, lake_prb22, rahul_prb18, peng_prr20, peng_prr21, tomasi_NHSE_prl}.

Experimentally, such models have been realized in cold-atom~\cite{Guardado_Sanchez_2020,Kohlert_2023,Wadleigh_2023,Scherg2021} and trapped-ion~\cite{Morong2021} and supercunducting~\cite{Gou_2021} setups in the presence of a tilted field of strength $\Delta$. In the limit of a strong field, dipole conservation emerges, i.e., 
$
[H,\mathcal{D}] =0$, 
where $\mathcal{D}$ is the dipole moment defined as, 
$
\mathcal{D} = \sum_j j n_j,
$
$n_j$ is the charge density, and the effective model exhibits Hilbert-space fragmentation, meaning that in the Krylov basis the Hilbert space splits into exponentially many dynamically disconnected blocks, resulting in partially or completely avoided thermalization.

\begin{figure}[!tb]  
     \includegraphics[width=1.0\columnwidth]{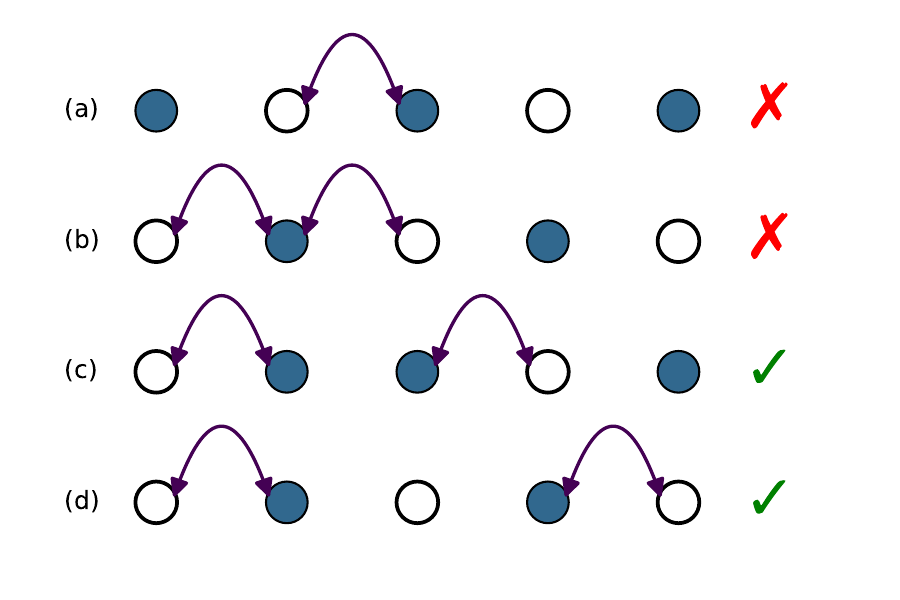}
    \caption{A schematic of the possible exchange of spins is shown here. Up spins are represented by filled circles, whereas down spins are shown as open circles. The spin exchange in (a) is not allowed because it involves the single spin exchange, and  (b) is not allowed because it requires two spins of the same kind on the same site. Also, both (a) and (b) do not respect the dipole conservation. Only (c) and (d) respect the dipole conservation and represent the $J$ term and $\gamma$ term of our Hamiltonian given by Eq.~\eqref{eq:ham}.}
    \label{fig:lattice}
\end{figure}

Experiments observe anomalous charge transport consistent with dipole conservation~\cite{exp1,exp2,Scherg2021,weitenberg_prx22}.
While this behavior can be understood perturbatively in the regime $\Delta \gg U, J$, this description breaks down for $\Delta \lesssim U$, where $U$ and $J$ denote the interaction and hopping strength, respectively. Nevertheless, anomalous dynamics persist for finite $\Delta > 0$ on experimentally accessible timescales. Since dipole conservation is only emergent, thermalization is ultimately restored at long times.

At low energies, at the level of the ground state, several models with exact dipole moment $\mathcal{D}$ conservation have been investigated. One example is the dipolar Bose-Hubbard model in one spatial dimension~\cite{BH_prb1, lake_prb22, knap_prb_23, lake_prl, kanp_BH}. Due to the interplay between the nonlinear constraints intrinsic to dipole conservation and the Hubbard interactions, these models exhibit a rich range of phases, including a fractonic Mott insulator, a dipole Luttinger liquid, and, in finite-size systems, a supersolid phase. The Mott phase is characterized by finite gaps in both the charge and dipole sectors. In contrast, the dipole Luttinger liquid is gapless in the dipole with algebraically decaying dipole correlations, while the charge sector remains gapped.

In this work, we investigate a paradigmatic one-dimensional spin-$1/2$ model having dipole constraints. We analyze magnetic ordering across different parameter regimes in the presence of an additional Ising interaction $U$, allowing for both ferromagnetic and antiferromagnetic couplings. To enforce dipole conservation, we employ the simplest pair-hopping Hamiltonian as shown in Fig.~\ref{fig:lattice}, which necessarily involves four spin operators. 

For short-range hopping-pair Fig.~\ref{fig:lattice}~(c), the model exhibits strong Hilbert space fragmentation and consequently fails to thermalize, meaning that the Hilbert space splits into exponentially many disconnected sectors, and where the number of states in the largest sector is still a vanishing fraction of the full Hilbert space dimension~\cite{hsf_1, hsf_2,hsf3,  gdt_2019}.

Furthermore, we introduce longer-range pair hopping Fig.~\ref{fig:lattice}~(d), controlled by a parameter $\gamma$, which induces spin frustration and breaks the sublattice symmetry. The inclusion of longer-range processes has important consequences for the Hilbert-space structure, driving a transition from strong to weak Hilbert-space fragmentation. In the weakly fragmented regime, the Hilbert space still splits into exponentially many sectors; however, the largest sector scales proportionally to the full Hilbert space at fixed macroscopic quantities, e.g., the magnetization and dipole moment~\cite{Moudgalya_RP22}.

In this model, we analyze the interplay between $U$ and $\gamma$, map out the phase diagram, and show that second-neighbor pair hopping not only alters the structure of the connected Hilbert (Krylov) subspaces, but also gives rise to additional ground state phases.

To characterize the nature of the ground state, we use both single-spin and dipole or doublon correlations. A doublon denotes a composite two-site object formed by correlations between a pair of spins. The single spin correlations are described by two-point correlation functions, which capture conventional magnetic ordering of individual spins. In contrast, doublon correlations are described by four-point correlation functions defined on nearest-neighbor pairs at different positions.

We find that, despite strong kinetic constraints, dipolar magnetic order can still emerge. For example, in the nonfrustrated and non-Ising limit, $\gamma = 0$ and $U = 0$, the model can be mapped to a noninteracting XX chain defined on an effective lattice of size $L/2$. For finite interaction $U > 0$, the model maps onto an XXZ chain with dipolar degrees of freedom instead of spins. This effective model exhibits the usual gapless Luttinger liquid to gapped transition at $U = 2 J$.

\begin{figure}[!tb]  
     \includegraphics[width=1.0\columnwidth]{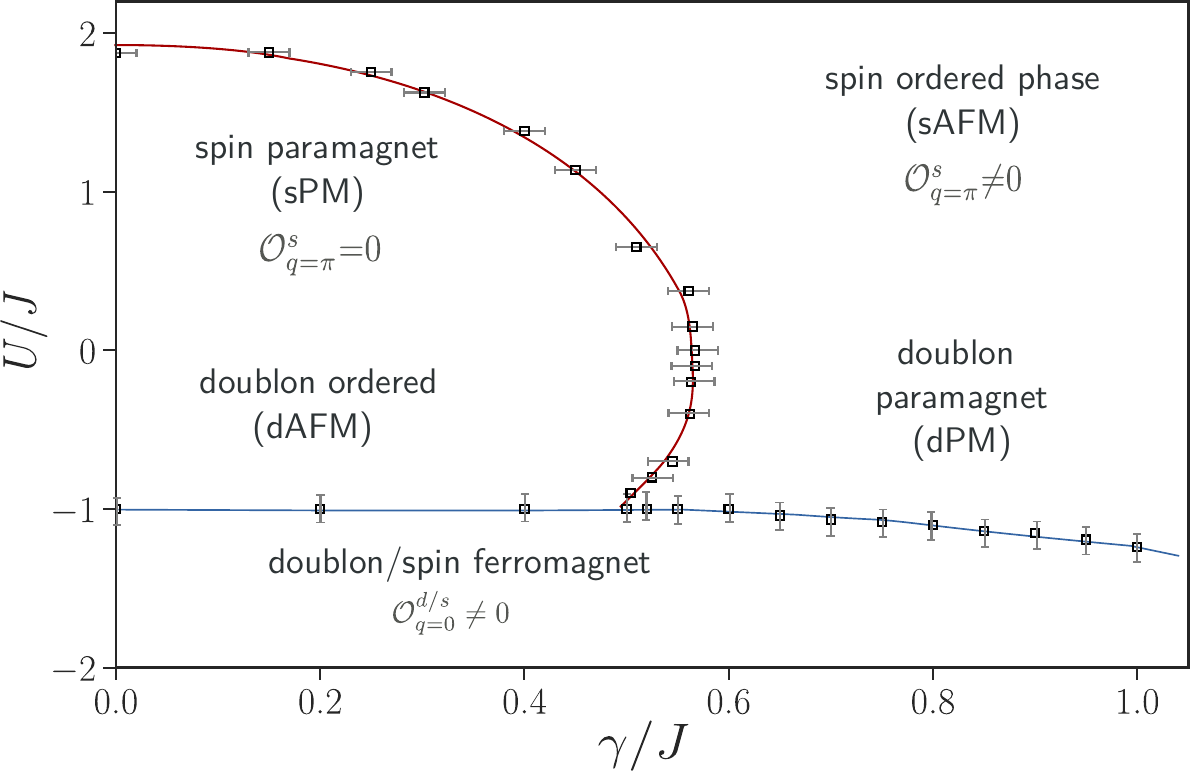}
    \caption{Finite size phase diagram illustrating the transition between the paramagnetic, antiferromagnetic, and ferromagnetic phases of spin and doublon. The phase boundary is identified by four independent indicators: the static structure factor at $q=\pi$, the doublon static structure factor at $q=\pi$, the dipole moment, and the entanglement spectrum. Results obtained using DMRG for $U\ge0$  with $L=128$,\, $\chi=384$, and for $U<0$, diagonalization is used with system size $L=28$.
    }
    \label{fig:pd}
\end{figure}

In the presence of finite frustration $\gamma$, such a mapping is no longer possible, as the sublattice symmetry is broken. We therefore resort to numerical calculations using the density matrix renormalization group method. Our results show that finite frustration can drive a transition from a doublon-ordered phase (dAFM) to a paramagnetic phase (dPM)~(see Fig.~\ref{fig:pd}).

We further investigate the dynamical response by computing the frequency-dependent spin susceptibility, which probes the excitation spectrum. In the magnetically ordered phase, the low-energy spectral weight is consistent with the formation of antiferromagnetic order. 
Additionally, the ground-state entanglement entropy and entanglement spectrum provide complementary information, identifying the phase boundary, including a non-divergent behavior at the phase transition, unlike the XXZ model. 
Taken together, the ground state correlations, dynamical response, and entanglement structure provide a coherent picture of the strongly correlated phases stabilized by dipole conservation in this model~\eqref{eq:ham}.

The paper is organized as follows. In Sec.\ref{model}, we introduce the model in detail and discuss its key symmetries, the construction of the Krylov subspace, and observables investigated in this study. We then present the phase diagram and provide an in-depth analysis of the different regimes in Sec.\ref{results} and conclude in Sec.\ref{discuss} with an outlook in Sec.\ref{outlook}.

\section{Model}
\label{model} 
\par We consider a dipole conserving a one-dimensional spin-1/2 chain of L sites, incorporating both multi-spin exchange interactions and nearest-neighbor Ising-type interactions. The model Hamiltonian of systems is given as,
\begin{align}
H& = {J}\sum_{i=1}^{L-3} \left( S^{+}_{i}S^{-}_{i+1}S^{-}_{i+2}S^{+}_{i+3} + h.c.\right)\nonumber \\
&+\gamma\sum_{i=1}^{L-4} \left( S^{+}_{i}S^{-}_{i+1}S^{-}_{i+3}S^{+}_{i+4} + h.c.\right)+{U}\sum_{i=1}^{L-1}S_{i}^{z}S_{i+1}^{z} .
\label{eq:ham}
\end{align}
 here, $S_{i}^{+}$, $S_{i}^{+}$, and $S_{i}^{z}$ are the usual spin raising, lowering, and $z$ component of the spin operator. The $J$ represents four-spin correlated exchange process($..\downarrow \uparrow \uparrow \downarrow... \rightarrow ..\uparrow \downarrow \downarrow \uparrow...$) and  $\gamma$, represents a long range spin exchange ($..\downarrow \uparrow  \downarrow \uparrow \downarrow... \rightarrow ..\uparrow \downarrow  \downarrow \downarrow \uparrow...$) inducing frustration. A schematic of these terms is shown in Fig.~\ref{fig:lattice}. $U$ is the nearest-neighbor Ising interactions between the $z$ components of the spin operators. This two-spin movement arises directly from the leading-order effective description for a tilted chain with nearest-neighbor interactions~\cite{exp2, Scherg2021, Moudgalya}.
 
\subsection{Symmetries of $H$}
In spin language, the total $U(1)$ charge (magnetization) and the dipole moment operator are translated as,

\begin{equation}
S^z_{T}= \sum_i S^z_i,\qquad  \mathcal{D}=\sum_i i\,S^z_i.
\end{equation}

Both operators are conserved, and therefore
\begin{equation}
[H,S^z_{T}] = 0, \qquad [H,\mathcal{D}] = 0.
\end{equation}
This can be rephrased in terms of a symmetry transformation,
\begin{equation}
\begin{split}
    S_i^{+} &\rightarrow e^{- \mathrm{i} \alpha i} \, S_i^{+}, \\
    S_i^{-} &\rightarrow e^{ \mathrm{i} \alpha i} \, S_i^{-}, \\
    S_i^{z} &\rightarrow S_i^{z},
\end{split}
\end{equation}
which leaves the Hamiltonian invariant $\forall \alpha$.

Moreover, the model also have the discrete symmetry $\mathbb{Z}_2$, 
\begin{equation}
\begin{split}
    S_i^{\pm} &\rightarrow  \, S_i^{\pm}, \\
    %S_i^{-} &\rightarrow  \, S_i^{-}, \\
    S_i^{z} &\rightarrow \,-S_i^{z}.
\end{split}
\end{equation}

\subsection{Discussion on limits and mapping}

When both $\gamma$ and $U$ are zero, the Hamiltonian reduces to simple pair-flip exchange terms, which exhibit an additional sublattice occupation symmetry beyond the $U(1)$ and dipole conservation laws. In this limit, the system is strongly fragmented, and in the sector with $S_T^z = \mathcal{D} = 0$, the largest connected subspace has dimension $\binom{L/2}{L/4}$~\cite{Moudgalya}. This sector can be mapped to an effective XX spin chain with $L/2$ sites~\cite{hsf_1, Morong2021}. In this reduced representation, each effective site corresponds to a two-site configuration of the original chain, and each configuration contains $L/4$ up and $L/4$ down spins.

We define the mapping
\begin{equation}
\begin{split}
|\uparrow_{2i-1}, \downarrow_{2i} \rangle &\rightarrow |\Uparrow\rangle_i, \\
|\downarrow_{2i-1}, \uparrow_{2i} \rangle &\rightarrow |\Downarrow\rangle_i.
\end{split}
\end{equation}

The Hamiltonian then maps as
\begin{equation}
\sum_{i=1}^{L-3} \left( S_i^+ S_{i+1}^- S_{i+2}^- S_{i+3}^+ + \text{h.c.} \right)
\rightarrow
\sum_{i=1}^{L/2-1} \left( \tau_i^{+}\tau_{i+1}^{-} + \text{h.c.} \right),
\end{equation}

where $\tau_i^{-} |\Uparrow \rangle_i = |\Downarrow \rangle_i$, and the corresponding $z$-operator satisfies
$\tau_i^z |\Uparrow \rangle_i = \tfrac{1}{2}|\Uparrow \rangle_i$ and $\tau_i^z |\Downarrow \rangle_i = -\tfrac{1}{2}|\Downarrow \rangle_i$, satisfying Pauli commutation relations.

Upon turning on the interaction strength $U$, the Hamiltonian continues to conserve total charge, dipole moment, and sublattice occupation, and its Hilbert space fragmentation structure remains unchanged. In terms of the effective spins $|\Uparrow\rangle$ and $|\Downarrow\rangle$, the interaction term takes the form
\begin{equation}
U \sum_{i=1}^{L-1} S_i^z S_{i+1}^z
;\rightarrow
-U \sum_{i=1}^{L/2-1} \tau_i^z \tau_{i+1}^z-\frac{U L}{8}.
\end{equation}

This mapping changes the effective sign of the interaction: for example, a ferromagnetic coupling in the original representation becomes antiferromagnetic in the effective XX description, and vice versa.

As $\gamma$ is gradually turned on, it introduces frustration into the system. This frustration effectively breaks the sublattice symmetry, which is essential for mapping the Hamiltonian to an XXZ-type spin model. Consequently, the presence of $\gamma$ leads to a more complex interplay between $J$ and Ising interaction, which can not be explained within XXZ-type models.

\subsection{Krylov Sectors}
The presence of conserved quantities always leads to a block-diagonal structure of the Hamiltonian. As discussed, $H$ commutes with  $S^z_{T}$ and $\mD$; hence, the Hilbert space decomposes into Krylov subspaces labeled by the pair of quantum numbers $(S^z_{T},\mD)$:
$\mathcal{H}=\bigoplus_{S^z_{T},\mD}\mathcal{H}_{S^z_{T},\mD}$, where $\mathcal{H}_{S^z_{T},\mD}$ is the Hamiltonian of corresponding krylov subspaces. The ground state of the Hamiltonian will reside in one of the Krylov subspaces, represented by finite value of $(S^z_{T}, \mD)$.

It is important to note that when only $J$ is finite and $\gamma/J,U/J=0$, the ground state is unique and resides in the sector labeled by ($S^z_T = 0$,$\mD = 0$). This indicates that the system preserves both spin and dipole symmetries in its lowest-energy configuration. However, as $\gamma$ and $U$ are gradually turned on, we found that the ground state becomes degenerate and no longer lies entirely within the $\mD = 0$ subspace, but rather involves states with nonzero dipole moment, reflecting the emergence of a new phase induced by the finite $\gamma$ and $U$.

\subsection{Observables}
To characterize magnetic ordering, we analyze the two-spin correlation function, the static spin structure factor, which is defined at momentum $q$ as,
\begin{equation}
\mathcal{O}^{s}_{q} = \frac{1}{L^2}\sum_{i,j} e^{\iota q(i-j)}\langle S_i^z S_j^z \rangle.
\label{eq:sf}
\end{equation}
At $q=\pi$, $\mathcal{O}_{q}$ is a well-established order parameter for detecting antiferromagnetic ordering, as it captures staggered spin correlations across the lattice. The quantity $\os$ therefore provides a direct measure of antiferromagnetic correlations. For doublon/dipole ordering, we compute the four spin static structure factor, which is defined as,

\begin{equation}
\mathcal{O}^{d}_{q} = \frac{1}{L^2}\sum_{i,j} e^{\iota q(i-j)}\langle S_i^z S_{i+1}^z S_j^z S_{j+1}^z \rangle.
\label{eq:sfd}
\end{equation}
$\mathcal{O}^{d}_{q}$ at $q=\pi$ will be used to quantify the doublon ordering.

To further probe the nature of correlations in the ground state, we compute the spin-spin two-point correlation function, defined as
\begin{equation}
    C_s(r)=\langle S_1^z S_r^z \rangle .
    \label{eq:csr}
\end{equation}
and the four-point correlation function
\begin{equation}
    C_d(r)=\langle S^{z}_{1}S^{2}_{1}S^{z}_{r}S^{z}_{r+1} \rangle .
    \label{eq:cdr}
\end{equation}
These correlators characterize the spatial decay of magnetic correlations, providing information about the presence or absence of long-range magnetic ordering. Together, these quantities give a detailed picture of the magnetic ordering in the ground state.

Further, we define the dipole operator 
\begin{equation}
    d^{+}_{i}=S^{+}_{i}S^{-}_{i+1}, 
\end{equation}
to study the propagation of dipoles and compute the connected dipole correlation function,

\begin{align}
    \la d^{+}_{x}d_{x+r}\ra_\text{c} = \la S^{+}_{x}S^{-}_{x+1}S^{+}_{x+r+1}S^{-}_{x+r}\ra\\\notag
    -\la S^{+}_{x}S^{-}_{x+1}\ra\la S^{+}_{x+r+1}S^{-}_{x+r}\ra .  
\end{align}

The corresponding dipole density operator is defined as
\begin{equation}
   \la d^{+}_rd_{r}\ra_\text{c}= \la S^{+}_{r}S^{-}_{r+1}S^{+}_{r+1}S^{-}_{r}\ra-\la S^{+}_{r}S^{-}_{r+1}\ra\la S^{+}_{r+1}S^{-}_{r}\ra.
\end{equation}

Here, $\la  \cdot\ra_\text{c}$ refers to the connected correlation function. 
Throughout this work, the terms doublon and dipole are used interchangeably to refer to the same phase.

\begin{figure*}[!tbh]  
     \includegraphics[width=0.8\textwidth]{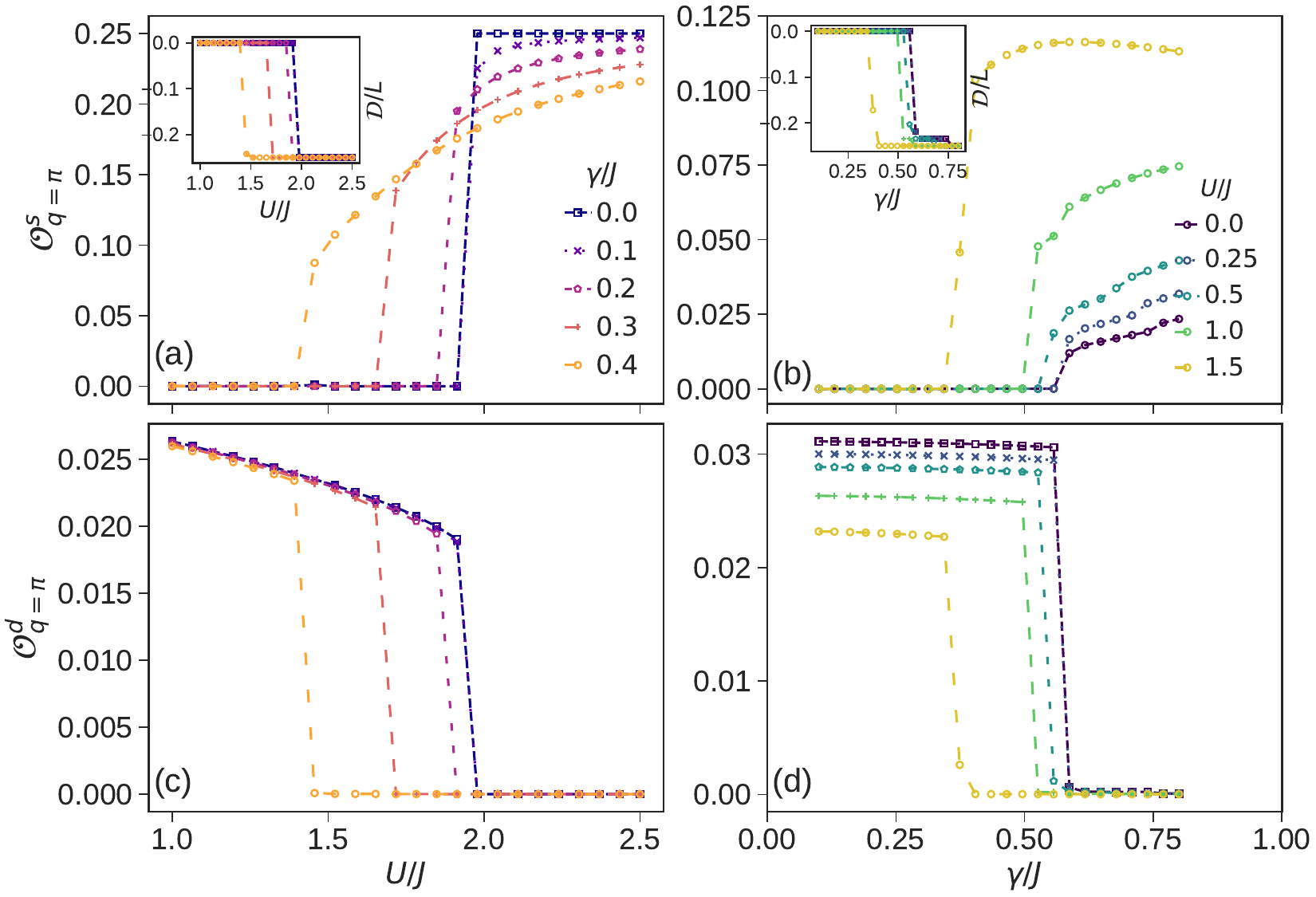}
    \caption{The static structure factor $\os$ is shown as a function of the interaction strength $U$ for  frustration strengths of $\gamma/J = 0.0, 0.1, 0.2, 0.3$, and $0.4$. The corresponding dipole moment is shown in the inset of panel (a). Additionally, the dependence of $\os$ on $\gamma$ is shown for various interaction strengths $U/J = 0.0, 0.25, 0.5, 1.0$, and $1.5$, and the inset of panel(b)  shows the corresponding dipole moment. The panel (c) illustrates the evolution of the four-spin static structure factor $\od$  as the interaction strength $U$ at fixed $\gamma$. Conversely, the variation of $\od$ with the $\gamma$ is shown for different interaction strengths in panel(d). The DMRG simulations are performed for a system size of $L = 128$ and a bond dimension $\chi = 384$.
    }
    \label{fig:order}
\end{figure*}

\subsection{Methods}
The density matrix renormalization group method (DMRG)~\cite{dmrg_prl1992, dmrg_aop2011} and Lanczos diagonalization~\cite{Lanczos1950,Elbio1994} have been widely used to study the ground state properties in quasi-one-dimensional systems~\cite{Giamarchi2004,cazalilla2011}. 
We have implemented DMRG simulation using the {\tt TeNPy} package \cite{dmrgTenpy}. Simulations are performed for a sufficient number of sweeps (min:50, max:200) and energy convergence is checked with a tolerance of $10^{-10}$. A mixer is used in the initial sweeps to improve stability and avoid local minima, with a small amplitude (${\sim} 10^{-3}$) that decays and is disabled after $10$ sweeps. We have used bond dimension, $\chi$, up to $1024$. Final results are reported only when energies are well converged. 

In the Lanczos method, we kept up to $50$ Lanczos vectors and maintained the orthonormalization accuracy of $\sim 10^{-12}$. In both DMRG and Lanczos simulations, we add a local magnetic field at the boundary site of the chain to lift the degeneracy. The field is kept sufficiently weak and does not affect the bulk properties of the system. It only serves to select a unique ground state and ensure numerical stability.

A comparison between the DMRG results and exact diagonalization is presented in Appendix~\ref{app:b}. The local-field dependence is discussed in Appendix~\ref{app:c}, while the system-size and bond-dimension dependences are examined in Appendix~\ref{app:d}.

We note that for antiferromagnetic interactions ($U/J > 0$), the DMRG method reliably converges to the ground state. In contrast, for ferromagnetic interactions $U/J < 0$, the ground state becomes degenerate, which makes convergence more difficult. In that regime, we therefore resort to Lanczos diagonalization for system sizes up to $L = 28$ to characterize the phase.

\section{Results}
\label{results}
%%%%
\subsection{Correlation functions}
%%%%
In this section, we study the ground-state phase diagram of the Hamiltonian in Eq.~\eqref{eq:ham} using DMRG~\cite{dmrg_prl1992,dmrg_prb1993,dmrg_aop2011,dmrgTenpy} and the Lanczos diagonalization method. We fix $J=1$, with all other parameters expressed in units of $J$. To map out the magnetic phases, we compute the static structure factor in Eq.~\eqref{eq:sf} and the four-spin structure factor defined in Eq.~\eqref{eq:sfd}. Figure~\ref{fig:pd} shows the resulting phase diagram of the dipole-conserving model as a function of $\gamma/J$ and $U/J$. We first discuss the limiting cases and then consider the full Hamiltonian. We will discuss separating the case in which the Ising interaction is anti-ferromagnetic ($U>0$) from the ferromagnetic ($U<0$).

\subsubsection{$\gamma/J=0$ $\&$ $U/J>0$}

In the limit $\gamma/J = 0$, when the frustration term is absent, the model exhibits a transition at $U/J \approx 2.0$ from a spin paramagnet (sPM) to an antiferromagnet (sAFM). The order parameter $\os$ at $\gamma=0$ is shown in Fig.~\ref{fig:order}(a), clearly indicating this transition. The sPM-sAFM transition is further confirmed by the dipole moment, which abruptly jumps to $-0.25$ at the transition point, as shown in the inset of Fig.~\ref{fig:order}(a). The value $\mD/L = -0.25$ corresponds to the dipole sector in which the Neel state is one of the basis states. This transition is driven by the competition between the pair-flip and Ising interactions: for $U/J < 2.0$, the pair-flip term dominates, enhancing spin fluctuations that suppress magnetic ordering.
 As $U/J$ increases, the Ising interaction becomes energetically favorable and stabilizes antiparallel alignment between neighboring spins, thus driving the emergence of the AFM phase. In contrast, when $U/J<2$ is small, dipole conservation allows dipole movement and gives rise to antiferromagnetic ordering of doublons, as depicted by the order parameter $\od$ in Fig.~\ref{fig:order}(b), which shows the antiferromagnet(dAFM) to doublon paramagnet (dPM) transition.

As discussed above, in this limit the block structure of the Hamiltonian maps onto a ferromagnetic XXZ chain,
\begin{equation}
\label{eq:H_block}
\begin{split}
    H_{\text{block}}  = & \sum_{i=1}^{L/2-1} \left( \tau_i^{+}\tau_{i+1}^{-} + \text{h.c.} \right)
    -U \sum_{i=1}^{L/2-1} \tau_i^z \tau_{i+1}^z
    - \frac{U L}{8} \\
    = & 2\sum_{i=1}^{L/2-1} \left( \tau_i^{x}\tau_{i+1}^{x} + \tau_i^{y}\tau_{i+1}^{y}
    -\frac{U}{2}\, \tau_i^z \tau_{i+1}^z \right)
    - \frac{U L}{8}.
\end{split}
\end{equation}

In this form, the model corresponds to an XXZ chain with anisotropy $\Delta = -U/2$. The Hamiltonian $H_{\text{block}}$ exhibits a paramagnetic-to-ferromagnetic $\mathbb{Z}_2$ symmetry-breaking transition at $U=2$ (i.e., $\Delta=-1$). For $U>2$, the ground space is spanned by the fully polarized states $
\left|\Uparrow\right\rangle^{\otimes \frac{L}{2}}, 
\left|\Downarrow\right\rangle^{\otimes \frac{L}{2}}$.

In terms of the original physical spins, these states correspond to the two Neel configurations,
\[
\left|\Uparrow\right\rangle^{\otimes \frac{L}{2}} = |\uparrow,\downarrow\rangle^{\otimes \frac{L}{2}}, \qquad
\left|\Downarrow\right\rangle^{\otimes \frac{L}{2}} = |\downarrow,\uparrow\rangle^{\otimes \frac{L}{2}},
\]
where each $\tau_i$ represents a two-site unit cell.

Numerically, we find that for $0 < U < 2$ the ground state lies in the sector with $M=\mathcal{D}=0$, which maps to $H_{\text{block}}$. This observation is further confirmed by the condition $d^{\dagger}_{2i} d_{2i}=0$ throughout this regime, as will be discussed later. Within this sector, we can therefore use known results for the XXZ chain to compute the ground-state energy.

We parametrize the anisotropy as
\begin{equation}
\Delta = \cos \gamma, \qquad \gamma \in \left(\frac{\pi}{2}, \pi\right),
\end{equation}
corresponding to $-1 < \Delta < 0$.

The ground-state energy density in the thermodynamic limit is given by the Bethe ansatz expression~\cite{Giamarchi_2003,Korepin_Bogoliubov_Izergin_1993}
\begin{equation}
\label{eq:e0}
e_\infty(\gamma)
= \left[
\frac{\cos\gamma}{4}
- \frac{\sin\gamma}{2\pi}
\int_{-\infty}^{\infty}
\frac{\sinh\big((\pi-\gamma)x\big)}
{\sinh(\pi x)\cosh(\gamma x)}\,dx
\right].
\end{equation}

Thus, for $0 < U < 2$, the ground-state energy per site is
\begin{equation}
\frac{E_0(L,U)}{L}
\simeq 
e_\infty\!\left(\Delta=-\frac{U}{2}\right)
- \frac{U}{8}.
\end{equation}
At $U=2$, we have
\[
\frac{E_0(L,U=2)}{L} = -\frac{1}{4} - \frac{1}{4} = -\frac{1}{2},
\]
which corresponds to the energy density of the two Néel states. Thus, for $U>2$, the ground states are the two Néel states, which are exact eigenstates of the Hamiltonian with energy density
\begin{equation}
\frac{E_0(L,U)}{L} = -\frac{U}{4}.
\end{equation}

This implies a level crossing at $U=2$, consistent with the closing of the excitation gap at the transition point. The critical point therefore matches the ferromagnetic transition of the XXZ chain at $\Delta = -1$, which has a Kosterlitz--Thouless nature.

\subsubsection{$U/J$=0 limit}
Now, we set the Ising-type interaction to zero and vary the parameter $\gamma$ in order to demonstrate the existence of a sPM-to-sAFM transition. The $J$ term favors nearest neighbor (NN) pair flips, while the $\gamma$ term facilitates the flipping of two spins separated by one site. As $\gamma$ increases, these pair-flip processes, separated by one site, become more prominent. This generates the spin defects in the lattice. Consequently, the system evolves from a disordered phase to the AFM phase, as shown by the order parameter $\os$ in the main of Fig.~\ref{fig:order}(c). An abrupt change in both $\os$ and $\mD$ at the transition point confirms the transition to the sAFM phase.  Furthermore, in the regime of small frustration strength($\gamma/J<0.58$), as expected, doublons exhibit antiferromagnetic ordering. When $\gamma/J\ge 0.58$, a system goes from dAFM to dPM phase as shown in  Fig.~\ref{fig:order}(d), which clearly shows that $\od$ remains finite when $\os$ is zero and vanishes at the transition point. Due to the extremely small value of the $\os$, we performed a finite-size scaling analysis to reliably characterize the behavior. A sharp change becomes evident upon increasing both the system size and the bond dimension in DMRG simulation, as shown in the Appendix~\ref{app:e}.

\subsubsection{$U/J > 0$, $\gamma/J > 0$}
In order to get a further understanding of the full phase diagram in the ($\gamma$, $U$) plane, we analyze four vertical and four horizontal cuts of the phase diagram. Main panel of Fig.~\ref{fig:order}(a) shows the order parameter $\os$ at fixed frustration strength by varying interaction strength across the transition point. We observe a sharp change in $\os$ near the transition point, and with an increase in $\gamma$, the transition point moves towards the lower $U$; hence, these exit the critical $U$ for each value of $\gamma$. This transition originates from frustration, which amplifies the Ising interaction's preference for antiferromagnetic alignment. As a result, the system develops a robust AFM order, which is further supported by the dipole moment in the inset of Fig.~\ref{fig:order}(a). In contrast, Fig.~\ref{fig:order}(b) depicts the order parameter $\od$, which shows that doublons are antiferromagnetically ordered precisely in the regime where spins are not ordered.

Furthermore, we show the order parameters and dipole moment for fixed interaction strength $U/J$ in Fig.~\ref{fig:order}(b). For a fixed $U$, there exists a critical value of $\gamma$ beyond which the spin order emerges. The order parameter $\os$ exhibits a sudden enhancement, indicating that the system undergoes a PM to AFM transition, which is corroborated by the dipole moment as shown in the inset of Fig.~\ref{fig:order}(c). As $U$ increases, a smaller value of $\gamma$ is sufficient to induce ordering, since a larger $U$ favors the antiparallel alignment of nearest-neighbor spins energetically. To examine the doublon order, we plot the $\od$ for the same parameters in Fig.~\ref{fig:order}(d). $\od$ grows in a range complementary to the spin-ordered regime, indicating that doublon order emerges only outside the spin-ordered regime. This complementary behavior highlights the mutually exclusive nature of the two orders.

\begin{figure}[!tb]  
     \includegraphics[width=1.0\columnwidth]{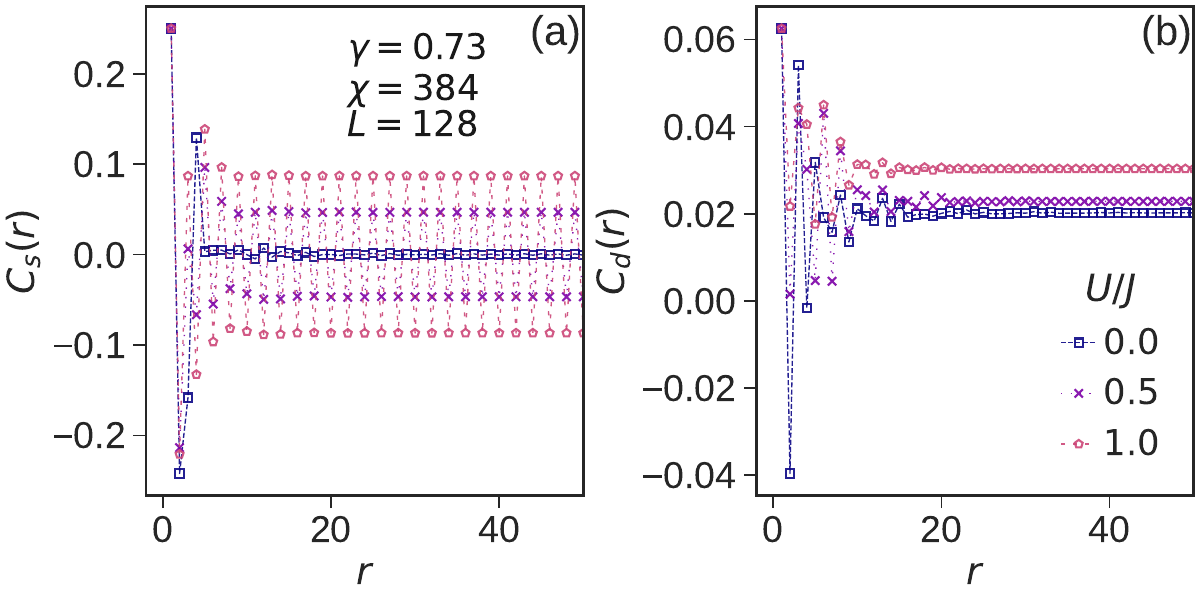}
    \caption{Spin-spin correlator   and doublon-doublon correlator  as a function of distance $r$ for $U/J=0.0,0.5,1.0$ in panel(a) and (b)respectively for fixed $L=128$ and $\chi=384$.}
    \label{fig:correls}
\end{figure}

We also studied the real-space spin and doublon correlators to confirm the presence of the ordered phase. Panel(a) of Fig.~\ref{fig:correls} depicts the behavior of the spin correlator $C_s(r)$, given by Eq.~\eqref{eq:csr}, for three representative values of $U/J$ for a fixed value of frustration strength $\gamma/J=0.73$. In the sAFM phase, the spin correlator exhibits a staggered pattern, and the orders become more prominent with increasing interaction strength, while $C_d(r)$ (Eq.~\eqref{eq:cdr}) shows the absence of a staggered pattern as shown in panel(b) of Fig.~\ref{fig:correls} and confirms the mutually exclusive nature of spin and doublon orders.

\begin{figure}[!tbh]  
     \includegraphics[width=1.0\columnwidth]{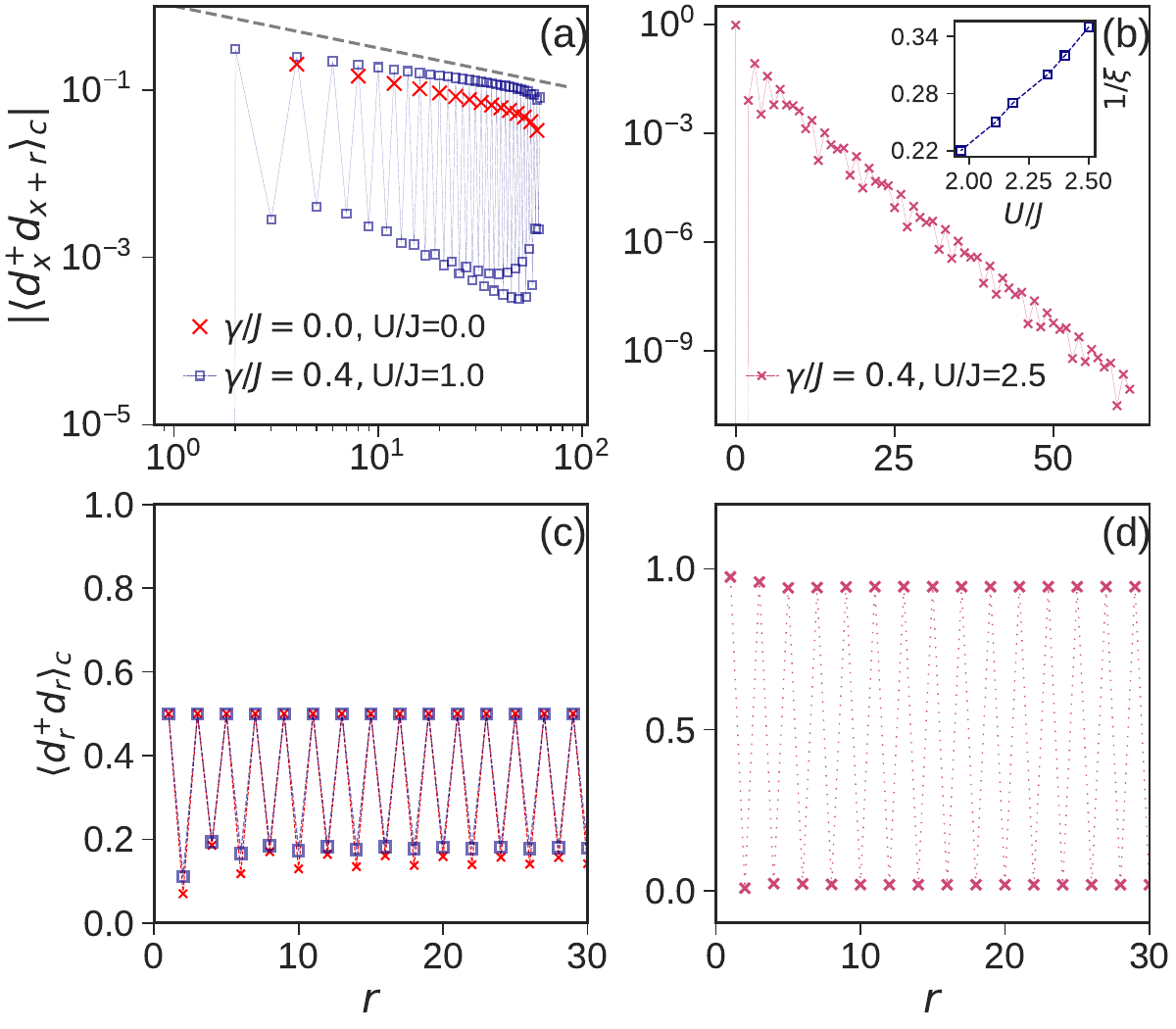}
    \caption{(a) The dipole correlator $\la d^{+}_xd_{x+r} \ra_{c}$ for $\gamma/J=0, U/J=0$ and $\gamma/J=0.4, U/J=1.0$ with $x=L/2$. The dashed line is a guide to the eye for the power-law decay exponent $1/2$. 
    (b) Shows the same data in the disordered phase $\gamma/J=0.4, U/J=2.5$, an exponential decay of the correlation is highlighted with log-y scale. Inset of panel(b) shows the correlation length dependence on interaction strength. The dipole density $\la d^{+}_{r}d_{r}\ra_{c}$ for the same parameters is shown in  (c) and (d). $L = 128$ and $\chi = 384$ remains fixed. 
    }
    \label{fig:dcorr}
\end{figure}

Finally, we computed the diagonal and off-diagonal dipole correlators. In the regime of doublon antiferromagnetism, the dipole correlator exhibits a power-law decay,
\[
|\langle d^{\dagger}_x d_{x+r} \rangle_{c}| \propto r^{-\alpha},
\quad \alpha = 1/2.
\]
In the dipole paramagnetic regime, it decays exponentially,
\[
|\langle d^{\dagger}_x d_{x+r} \rangle_{c}| \propto e^{-r/\xi},
\]

with correlation length $1/\xi = 0.35$ for $\gamma/J=0.4,\, U/J=2.5$, as shown in panels (a) and (b) of Fig.~\ref{fig:dcorr}, respectively.
With increasing \(U\), the correlation length \(\xi\) increases, as shown in the inset of Fig.~\ref{fig:dcorr}(b). To minimize boundary effects, the correlation length is determined by selecting a reference site at the center of the chain with $x=L/2$ and measuring correlations toward the edges.

For $\gamma/J = 0$, the exponent  \(\alpha = 1/2\)  at $U/J=0$ has a simple interpretation: in this regime, as already discussed, the ground state coincides with that of the block Hamiltonian in Eq.~\eqref{eq:H_block}, and one finds
\[
|\langle d^{\dagger}_x d_{x+r} \rangle_{c}|
\sim
|\langle \tau^+_x \tau^-_{x+r} \rangle_{c}|
\sim
\frac{\log r}{r^{1/2}}.
\]

For \(0<U<2\) and $\gamma = 0$, we expect instead
\[
|\langle d^{\dagger}_x d_{x+r} \rangle_{c}|
\sim
\frac{1}{r^{1/2K}},
\]
where the exponent is determined by the Luttinger liquid parameter $K = \frac{\pi}{2(\pi-\arccos{-\frac{U}{2}})}$~\cite{Giamarchi_2003}.

For $\gamma/J=0,U/J=0$, correlation at all even sites is zero and slowly develops with the introduction of finite interaction and frustration at even sites also. The diagonal dipole correlator $d^{+}_{r}d_{r}$  in the dipole AFM phase oscillates with r. Using the definition $d_i^\dagger=S_i^+S_{i+1}^-$, the dipole density $d_i^\dagger d_i $ probes nearest-neighbor spin configurations. By using the spin-1/2 identities $S^+S^-=1/2+S^z $ and $S^- S^+ =1/2 - S^z $, we will get $d_i^\dagger d_i=(1/2 + S_i^z)(1/2-S_{i+1}^z)$. This operator equals unity for $\uparrow_{i} \downarrow_{i+1}$ and vanishes for $\downarrow_{i} \uparrow_{i+1}$. In the  Neel AFM state, such bonds alternate between sites, which results in an oscillatory dipole density that becomes unity on odd sites and vanishes on even sites. This staggered pattern represents a frozen Neel configuration and indicates that the system is deep in the sAFM phase. In contrast, for the dAFM phase, the spins fluctuate quantum mechanically and are not locked into a fixed pattern. Hence, both factors in the expression $d_i^\dagger d_i=(1/2+S_i^z)(1/2-S_{i+1}^z)$ acquire finite expectation values on every bond, resulting in a nonzero value at all sites.

\subsubsection{$U/J<0$ : Ferromagnetic Interactions}
\begin{figure}[!tb]  
     \includegraphics[width=1.0\columnwidth]{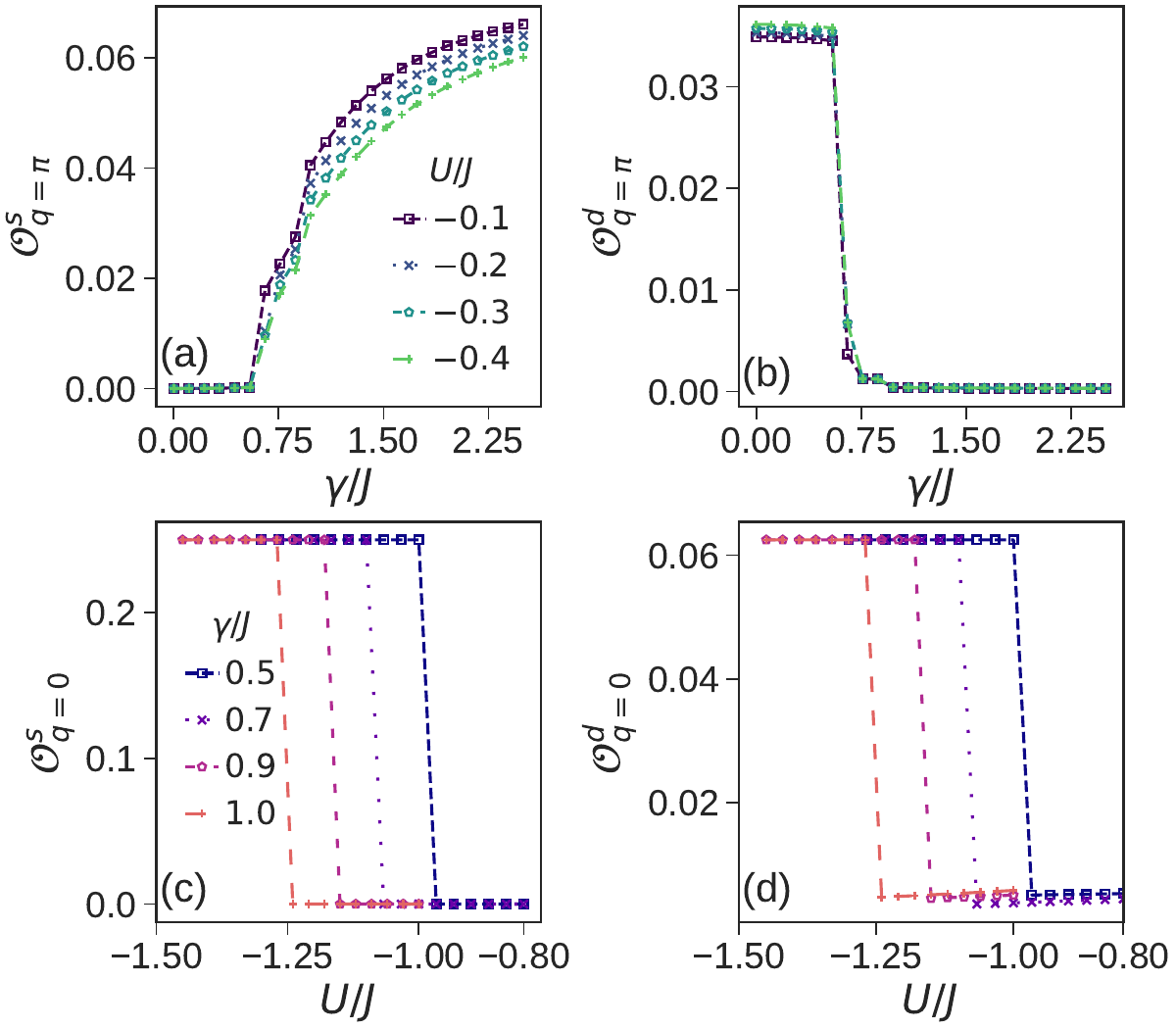}
    \caption{(a) The static structure factor $\os$ is shown as a function of the frustration strength $\gamma$ for $U/J = -0.1, -0.2, -0.3,-0.4$. Additionally, the dependence of $\od$ on $\gamma$ is presented for the same interaction strengths in panel(b). The panel (c) illustrates the evolution of the static structure factor $\osO$  as the interaction strength $U$ at fixed frustration $\gamma=0.5,0.7,0.9,1.0$ and doublon structure factor $\odO$ in panel(d). The simulations are performed for a system size of $L=28$ with the Lanczos method.}
    \label{fig:order_negativeU}
\end{figure}

We extended our analysis to study the interplay between ferromagnetic coupling and frustration strength. 
We first examine the role of frustration by analyzing the $\os$ and the doublon structure factor $\od$ at fixed interaction strengths $U$. Panel (a) and (b) of Fig.~\ref{fig:order_negativeU} dipict the $\os$ and $\od$ as a function of  $\gamma$ for fixed interaction strength. For each  $U$, there exists a critical $\gamma$ above which the system goes to the AFM phase, which can be seen in panel (a). At the same time, the double structure factor vanishes after that critical strength, as shown in panel (b), which demonstrates the transition from dAFM to sAFM phase.

When the stenght of the interaction is negative, it is natural to expect a ferromagnetic phase. We indeed found the spin and doublon ferromagnetic regimes. Figure~\ref{fig:order_negativeU}(c) shows the structure factor $\osO$ at ferromagnetic wave vector for four diffrent value of $\gamma/J$. For fix frustration strength, we found the critical interaction strength at which the system enters the FM phase, and as $\gamma/J$ increases, the critical $U/J$ decreases. The doublon structure factor is shown in panel (d) of Fig.~\ref{fig:order_negativeU}, which clearly shows the transition doublon FM phase. The coexistence of spin and doublon FM arises because the ground state in this parameter regime is fully polarized. All results for $U/J<0$ are obtained by performing the Lanczos diagonalisation for $L=28$.

For the limiting case $\gamma = 0$, the ferromagnetic $U$ regime is more subtle with respect to the mapping. In fact, if we restrict the Hamiltonian to the largest blocks that can be mapped onto the XXZ model, we find that a ferromagnetic phase cannot be realized within this effective description. It is sufficient to consider the case where $U$ negatice but large in absolute value: the Hamiltonian in Eq.~\ref{eq:H_block} approaches the two Neel states in the effective spins, $|\Uparrow, \Downarrow \rangle^{\otimes L/2}$ and $|\Downarrow, \Uparrow \rangle^{\otimes L/2}$, which in terms of the physical spins correspond to $|\uparrow, \downarrow, \downarrow, \uparrow \rangle^{\otimes L/4}$ and its $\mathbb{Z}_2$-transformed partner, both having a domain size equal to two. In other words, the mapping to the original spins in the XXZ picture does not allow one to host a ferromagnetic phase. 

To confirm the transition, we used the approximate expression for the ground-state energy Eq.~\eqref{eq:e0}, which is $e_\infty(U)\approx   -\frac{1}{\pi} + \frac{U}{2 \pi^2}$~\cite{Korepin_Bogoliubov_Izergin_1993}. In order to see the transition in the case, the following inequality
\begin{equation}
    e_\infty(U) -\frac{U}{8} < \frac{U}{4},
\end{equation}
must be violated, $U< \frac{8\pi}{4-3\pi^2} \approx -0.98$ as confirmed in Fig.~\ref{fig:pd}  and Appendix~\ref{app:f}.

\begin{figure}[!tbh]  
     \includegraphics[width=1.0\columnwidth]{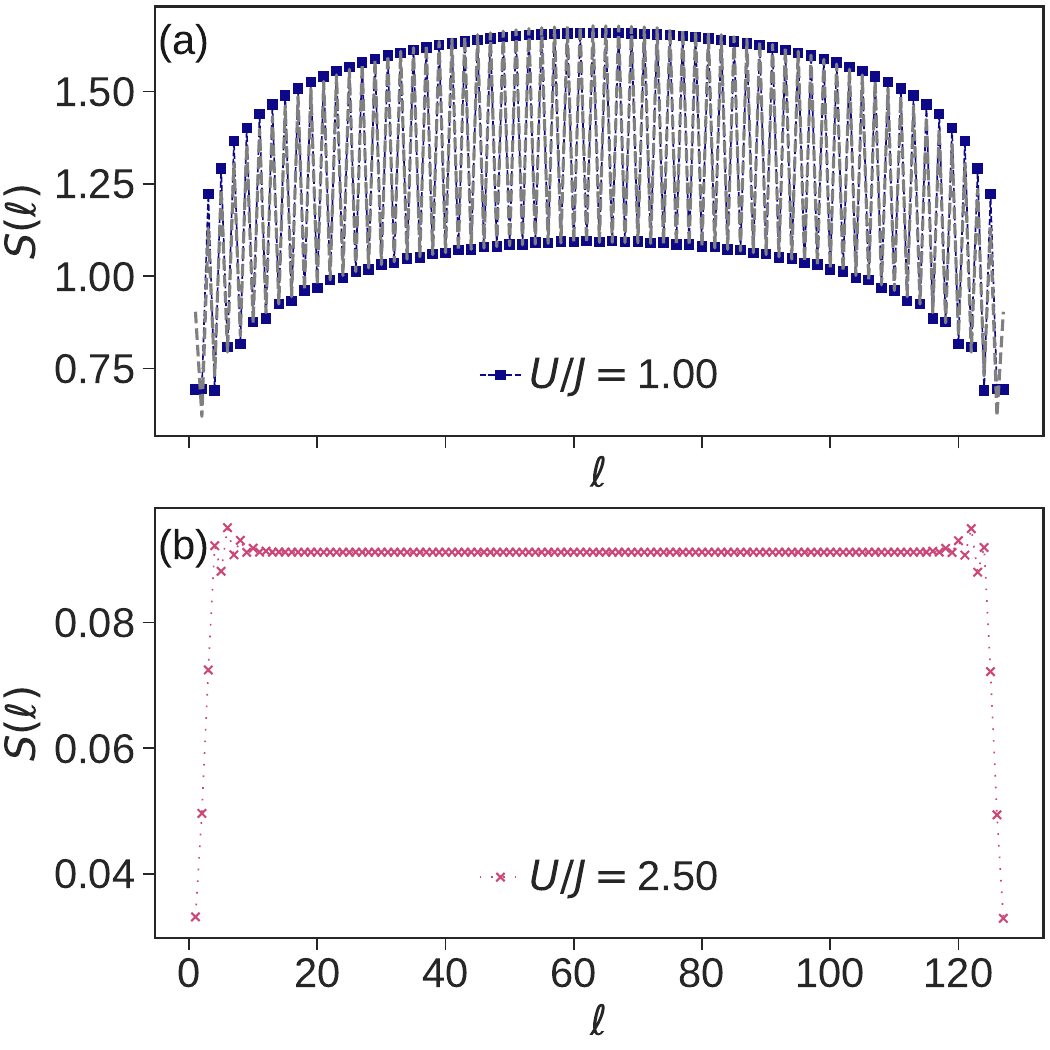}
    \caption{Entanglement entropy for fixed $\gamma/J=0.2, L=128, \chi=384$ and two value of $U$: one in doublon antiferomagnet $U/J=1.0$ in (a) and one in doublon paramagnet $U/J=2.5$ in (b). The central charge is obtained by fitting the EE using the Calabrese-Cardy formula. In the dAFM phase, $S(\ell)$ shows oscillatory behavior with $c=1$, while in the dPM regime, $c=0$.}
    \label{fig:EE}
\end{figure}

\subsection{Entanglemet Entropy}
Entanglement entropy (EE) is a measure of quantum correlations and quantifies the extent to which a subsystem is entangled with the rest of the system. In particular, the central charge, $c$, as extracted from EE scaling, identifies the underlying conformal field theory (CFT) and distinguishes gapless critical phases from gapped ones across the transition. 

We compute the central charge by employing the Calabrese-Cardy formula~\cite{Calabrese_2009,Fagotti_2011,affeleck_prl2006}, 
\begin{equation}
S(\ell)= \frac{c}{6}\,\log\!\left[\frac{2L}{\pi}\sin\!\left(\frac{\pi\ell}{L}\right)\right]+ A \,\frac{\cos(\pi \ell)}{\left[\frac{2L}{\pi}\sin\!\left(\frac{\pi\ell}{L}\right)\right]^{p}}. 
\end{equation}
where $l$ is the sub-system size, $c$ is the central charge, $A$ is the amplitude of oscillations, and $p$ is the decay exponent of oscillations, which depend on the interaction. In previous studies, EE has been used to characterize its critical behavior~\cite{korepin2004}. In the gapless Luttinger liquid phase, the EE scales logarithmically with subsystem size, consistent with CFT with central charge $c=1$. 

We have examined the EE at fixed coupling $\gamma/J=0.2$ for two interaction strengths corresponding to the dAFM phase at $U/J = 1.0$ and the dPM phase at $U/J = 2.5$ as shown in panels (a) and (b) of Fig.~\ref{fig:EE} respectively. In the dAFM phase, the EE exhibits oscillatory behavior as a function of the subsystem size $\ell$, reflecting the underlying dipolar order. The fit yields a central charge $ c\approx 1$, consistent with a phase described by a conformal field theory with $  c\approx 1$. This is closely analogous to that behavior observed in the dipolar Bose-Hubbard model~\cite{BH_prb1}, which supports the interpretation of the dipole antiferromagnetic phase. In contrast, in the dPM phase, the EE rapidly saturates, oscillations are absent, and the extracted central charge is strongly suppressed with $c\approx 0$. 

The magnitude of the entanglement entropy also distinguishes the phases. In the dAFM phase, the entropy remains relatively large, $\sim 2 \ln 2$, indicating a highly entangled state. In contrast, in the dPM or sAFM phases, it is much smaller, $\sim 0.1 \ln 2$, indicating that the state is weakly entangled and close to a product state.

\subsubsection{Entanglement Spectrum}
The entanglement spectrum (ES)  is routinely used \cite{ES_prl,ES_prb,ES_PRB2} to investigate phase transitions. 
ES is determined from the Schmidt decomposition of the ground state $|\phi\rangle $ and defined as, $|\phi\rangle = \sum_i \sqrt{\lambda_i}\, |\phi^A_i\rangle \otimes |\phi^B_i\rangle$. Here, $\lambda_j$ denotes the Schmidt coefficients, while $|\phi^A_i\rangle$ and $|\phi^B_i\rangle$ are the orthonormal basis for the subsystems $A$ and $B$, respectively. The reduced density matrix of subsystem $A$, defined as $\rho_A = \mathrm{Tr}_B(|\phi\rangle \langle \phi|)$, can be expressed in a thermal-like form, $\rho_A = \sum_i e^{-\varepsilon_i}\, |\phi^A_i\rangle \langle \phi^A_i|$, where the $\varepsilon_i$ are referred as entanglement energies. 

\begin{figure}[!tbh]  
     \includegraphics[width=1.0\columnwidth]{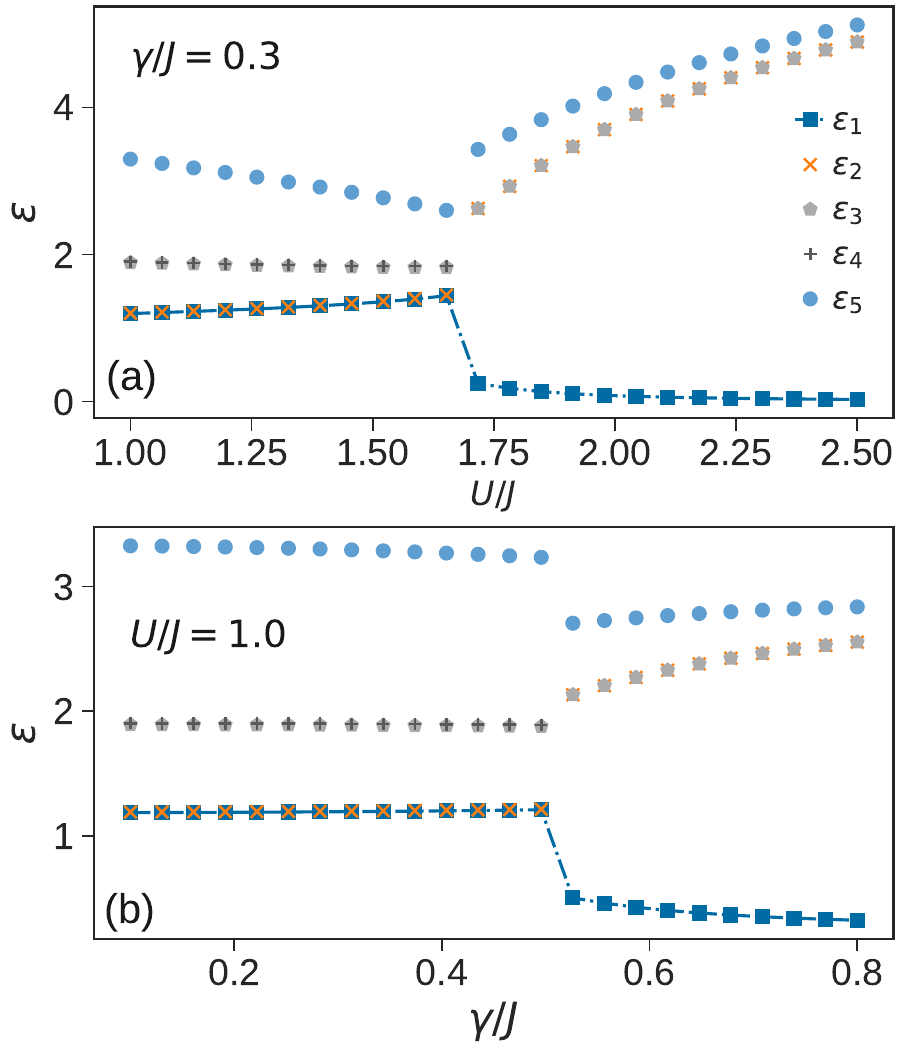}
    \caption{Entanglement spectrum of the ground state for $\gamma/J=0.3$  versus $U$ in panel (a), while panel (b) shows for $U/J=1.0$ as a function of $\gamma$.  In both panels, the first five entanglement energies are shown for $L=128$ and $\chi=384$. The lowest entanglement energy becomes non-degenerate at the transition point.}
    \label{fig:entspec}
\end{figure}

We show the five lowest levels of ground-state ES in Fig.~\ref{fig:entspec} for fixed $\gamma/J=0.3$  in panel(a) by changing the interaction strength. It is well known~\cite{ES_prb} that at the transition point, the lowest entanglement energy, $\varepsilon_{1}$, becomes non-degenerate. Indeed, $\varepsilon_{1}$ remains degenerate up to $U/J\approx 1.7$ and after that becomes non-degenerate, indicating transition to the antiferromagnetic phase. Panel (b) shows the ES at $U/J = 1.0$ as a function of frurster strength $\gamma$. The first entanglement energy, $\varepsilon_{1}$, becomes non-degenerate after the critical value of $U/J$, which shows the PM to AFM transition. In the XXZ chain~\cite{ES_prb}, the lowest entanglement level $\varepsilon_{1}$ becomes nondegenerate at the transition, while all higher entanglement energies diverge, leaving only $\varepsilon_{1}$ finite. In contrast, in our system, the higher entanglement energy levels do not diverge at the transition point.

\subsection{Dynamical Suceptibility}

We now focus on the dynamical susceptibility, which provides direct information about the collective excitations of the underlying system. We study the excitation spectra of the ground state of the dipole conserving model defined in Eq.~\eqref{eq:ham} by computing the dynamical susceptibility $S(q,\omega)$, which can be measured in Bragg spectroscopy~\cite{stenger1999} in optical lattices and inelastic neutron scattering experiments~\cite{ins} in solids, defined as,
\begin{equation}
S(q, \omega) = \sum_{n} \left| \langle \psi_{n} | O | \psi_{0} \rangle \right|^2 \delta\left( \omega - (E_n - E_0) \right),
\label{eq2:dF}
\end{equation}

where $O= \frac{1}{\sqrt{L}} \sum_{j=1}^{L} e^{-\iota q j} S^{z}_j$. Here $| \psi_{0} \rangle$ and $E_0$ are the ground state and energy of $H$, and the summation over $n$ runs over all the excited states of $H$. We demonstrate that the excitation spectra in both phases exhibit distinct types of excitation. We compute $S(q,\omega)$ using Lanczos diagonalization with a fixed system size of $L=28$.

\begin{figure}[!tb]  
     \includegraphics[width=1.0\columnwidth]{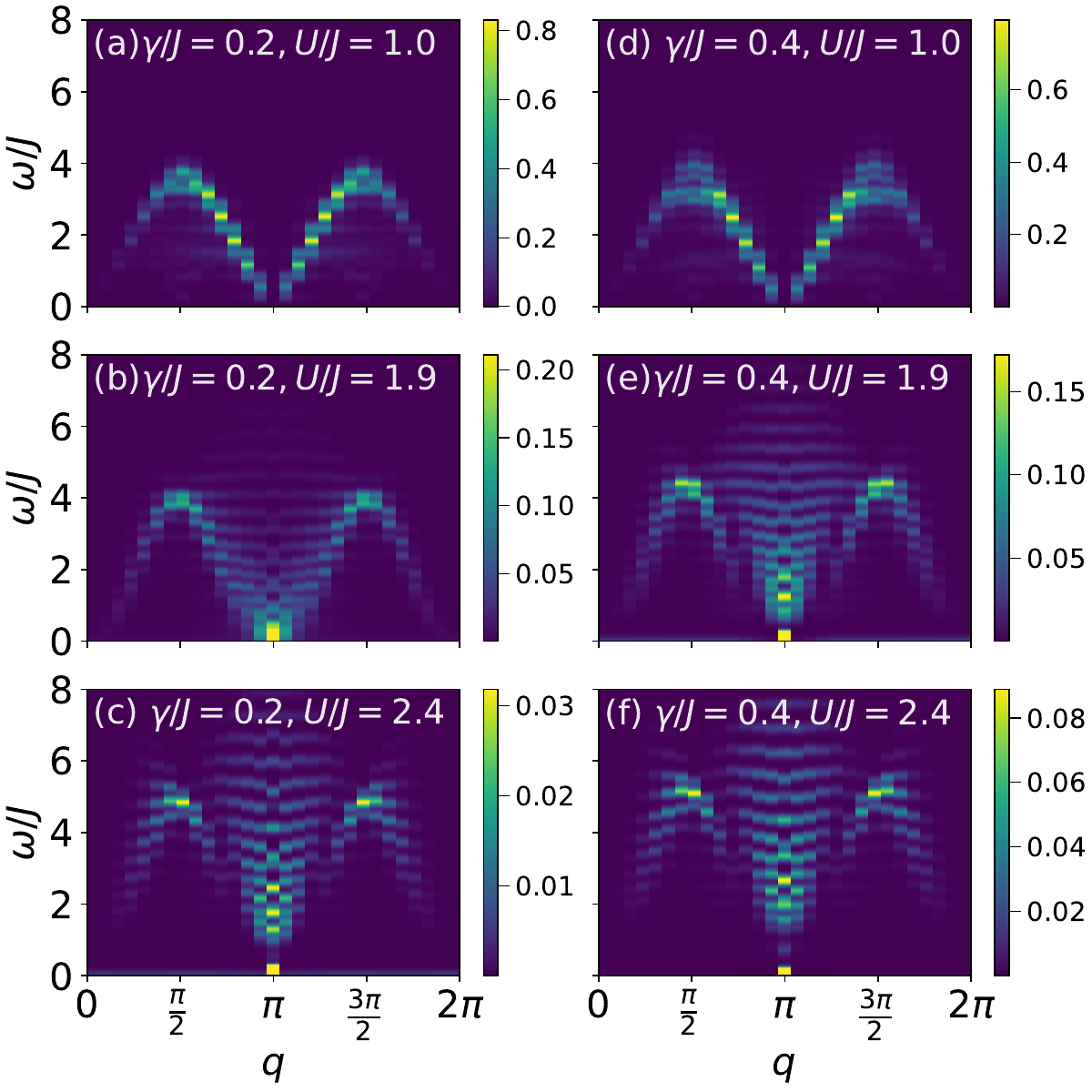}
    \caption{Dynamical spin structure factor $S(q,\omega)$ for fixed $\gamma/J$ and different interaction strength $U/J$. Panel $(a)-(c)$ corresponds to the $\gamma/J = 0.2$ and $U/J=1.0,1.9,2.4$, and $(d)-(f)$  $\gamma/J = 0.4$ and $U/J=1.0,1.9,2.4$. In the ordered phase, enhanced spectral weight appears near the antiferromagnetic wave vector $q=\pi$ and a clear separation of the two branches, indicating that the nature of the ground state is antiferromagnetic.}
    \label{fig:sucep}
\end{figure}

In Fig.~\ref{fig:sucep}, we present the dynamical spin structure factor  $S(q,\omega)$ at fixed frustration strength for three $U$ values: one in the deep AFM phase $U/J=2.4$, one near the transition point $U/J=1.9$, and in the disordered phase $U/J=1.0$. Figure~\ref{fig:sucep}$(a)-(c)$ shows the dynamical suceptibility for $\gamma/J = 0.2$. We found that the spectral response is relatively broad and featureless at wave vector $q=\pi$, with only a redistribution of weight across momentum and frequency in the paramagnetic regime. In particular, there is no pronounced signature of spinon continuum~\cite{BA1,BA2}, which are hallmarks of the AFM ground state, indicating that the system remains far from exhibiting well-developed staggered spin correlations. At sufficiently large $U$, the spectral weight begins to concentrate near the antiferromagnetic wave vector, and well-defined intensity peaks emerge at $q=\pi$. Panel $(d)-(f)$ correspond to $\gamma/J=0.4$ and showed simliar spectral distribution as for $\gamma/J=0.2$. Thus, the evolution of the spectral function with increasing interaction strength further provides evidence of the transition from a spin paramagnetic regime to one exhibiting the characteristic excitations of the AFM phase.

\section{Discussion}
\label{discuss}
The one-dimensional spin-$1/2$ model Eq.~\eqref{eq:ham}, with conserved $U(1)$ charge and dipole moment $\mD$, 
is of particular interest from the perspective of nonequilibrium dynamics. 
It provides a prototypical example of ergodicity breaking at infinite temperature due to kinetic constraints, which also shows strong Hilbert space fragmentation. For longer pair-hopping terms ($\gamma\ne0$), the sublattice symmetry is broken, and a weaker form of fragmentation persists, eventually leading to thermalization, as described by fractonic hydrodynamics~\cite{NandkishoreRev19,Grosvenor22}.

Here, we investigate the ground state of this model by studying the interplay between the frustration parameter $\gamma$ and the Ising interaction $U$.
We determine the numerical phase diagram and identify a transition from a doublon-antiferromagnetic phase to a spin-antiferromagnetic phase. The ordered doublon state remains stable up to critical values of the frustration and interaction strengths, beyond which the system reorganizes into a spin antiferromagnetic configuration. In the ferromagnetic regime $U < 0$, we observe a paramagnetic to antiferromagnetic transition as a function of frustration strength, and a paramagnetic to ferromagnetic transition upon varying the interaction strength.

A series of works charted the ground-state phase diagram of models with dipole $\mD$ constraints, such as the dipolar Bose-Hubbard model~\cite{lake_prb22, BH_prb1, knap_prb_23}. These models exhibit a dipole condensate phase in which the dipole correlator decays algebraically. 
The doublon AFM phase can indeed be compared with such a phase as the characteristic dipole correlator $|\la d^{+}_{x}d_{x+r}\ra_{c}| \sim r^{-\alpha}$ decays as a power law with a power $\alpha$. The exponent $\alpha$ is nonuniversal and is related to the dipole Luttinger parameter~\cite{BH_prb1, knap_prb_23}.
%{\tt please check notation}. 

In the doublon paramagnetic phase, $|\la d^{+}_{x}d_{x+r}\ra_{c}|\sim e^{-r/\xi}$ decays exponentially, consistent with the behavior expected in a dipole Mott phase. Unlike the spin model considered here, the dipolar Mott insulator in the bosonic realization exhibits the typical lobe structure in the phase diagram corresponding to different commensurate boson fillings.

In contrast to earlier works, we found a previously unexplored regime of strong ferromagnetic interactions, $U/J<-1$. Here, both the spin and the doublon develop ferromagnetic order simultaneously, leading to a phase in which spin and doublon ferromagnetism coexist and the ground state evolves into a fully polarised configuration. 

Typically, in one dimension, gapless phases with a single low energy mode are characterized by central charge $c = 1$, for example, the XXZ model in the regime $|\Delta| < 1$, where $\Delta$ denotes the $zz$ coupling. Such phases exhibit algebraic decay of correlations, which is characteristic of Luttinger liquid behavior. In our case, the dipolar antiferromagnetic phase also shows $c = 1$, consistent with a gapless dipolar Luttinger liquid in the presence of finite frustration $\gamma > 0$, while the charge sector remains gapped. 

In the bosonic dipolar model~\cite{BH_prb1, knap_prb_23}, it has been shown using Luttinger liquid theory that the transition from the dipolar antiferromagnetic phase to the disordered phase is governed by a Berezinskii Kosterlitz Thouless transition (BKT). Our numerical results are consistent with such a scenario~(see Fig.~\ref{fig:order_negativeU}(a)). However, due to limited precision in the DMRG data near the critical region, we are not able to conclusively establish the BKT nature of the transition in this work.

Overall, our results demonstrate that dipole-conservation constraints can stabilize unconventional ordered states at zero temperature, both in the presence and in the absence of frustration. Related phenomena associated with additional conservation laws have previously been discussed in other strongly constrained systems, such as quantum Hall liquids~\cite{SeidelPRL05}.

\section{Outlook}
\label{outlook}
To explore the robustness of the phases, future work can focus on more general settings. One can study the effects of longer-range interactions, disorder, and higher-dimensional lattices, where the interplay among interactions, lattice structure, and dipole conservation may lead to new dynamical regimes. Another possible extension of the present work could be the study of the ground state phases of the higher-moment conserving model. The conservation of higher moments can impose an even stronger constraint on the lattice, potentially leading to further fragmentation of Fock space, critical behavior, and fractionalisation in the excitation spectra.

Further, understanding magnetism in moment-conserving systems can lead to the design of constrained quantum systems and devices, where engineered conservation laws are harnessed to control coherence, information storage, and dynamical stability. Finally, it would be interesting to establish closer connections with experimentally relevant cold-atom platforms, such as tilted optical lattices, where dipole conservation is already realized \cite{exp2,Scherg2021} and predicted phases may be directly probed.

\section{Acknowledgment}
GDT thanks Jacopo Gliozzi for many illuminating discussions on related topics.
P would like to acknowledge the financial support from
IIT Bombay through the Institute Post-Doctoral Fellowship. SB and P thank the National Supercomputing Mission~(NSM) for providing computing resources of `PARAM Porul' at NIT Trichy, and of `PARAM Rudra' at IIT Bombay 
implemented by C-DAC and supported by the Ministry
of Electronics and Information Technology (MeitY) and
Department of Science and Technology~(DST), India. 

\section*{Appendix} 
\appendix

\section{Comparison between ED $\&$ DMRG}
\label{app:b}
\begin{figure}[!tbh]  
     \includegraphics[width=1.0\columnwidth]{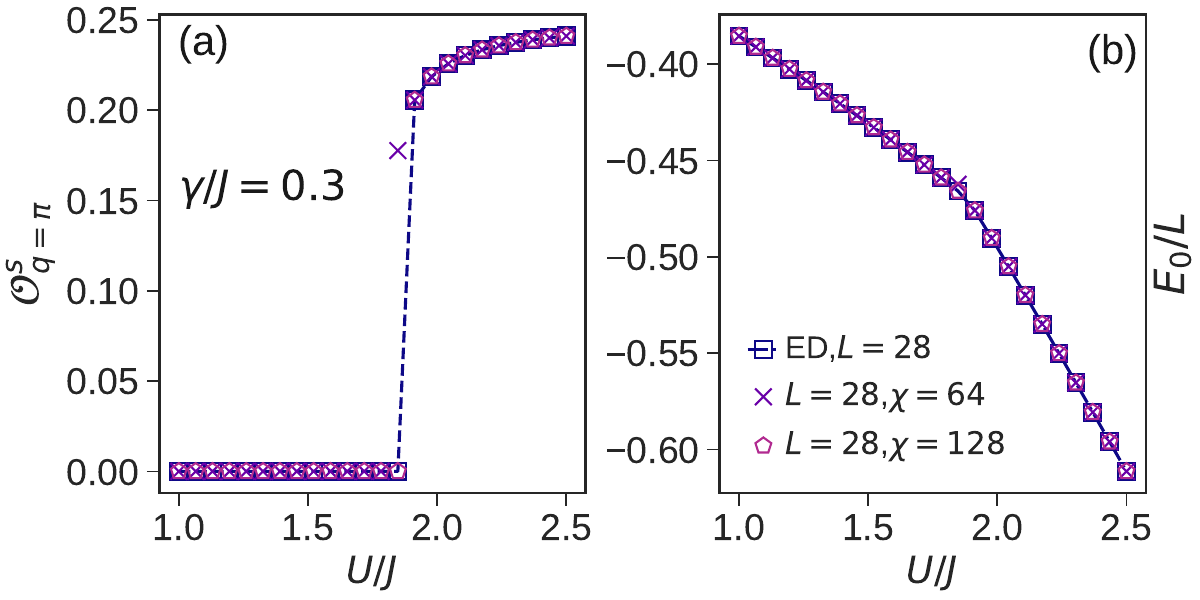}
    \caption{(a) The order parameter $\os$  using exact diagonalisation and DMRG simulation, while panel (b) shows the ground state energy for $L=28$ and two bond dimensions $\chi=64,128$.}
    \label{fig:spi_ed}
\end{figure}

We computed the static spin structure factor at $q=\pi$ using exact diagonalization for $L=28$, and our ED results indeed show a PM-to-AFM transition. Fig.~\ref{fig:spi_ed} shows the $\os$ for $\gamma=0.3$ in panel (a) and groundstate energy per system size in panel(b), respectively. We show the DMRG results for bond dimension $\chi=64$ and 128. Our ED and DMRG data agree very well. This concordance between results from ED and DMRG gives evidence that  PM to AFM transition reflects ground-state physics of the model rather than artifacts of system size or numerical method.

\begin{figure}[!tbh]  
     \includegraphics[width=1.0\columnwidth]{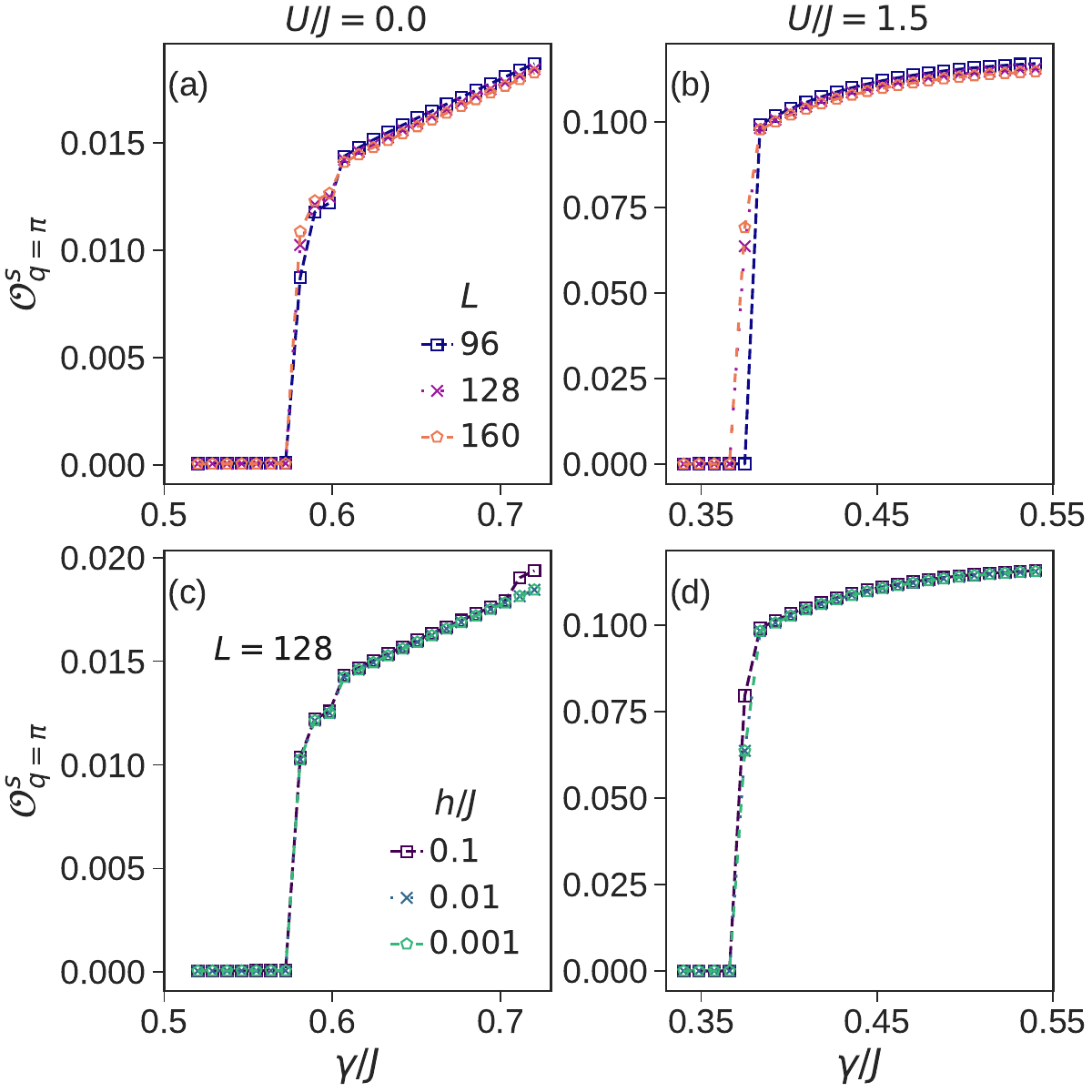}
    \caption{Panel (a) and (b) shows,$\os$ for $U=0.0$ and $U=1.5$ for $L=96,128,160$ while panel( c), (d) shows dependence of local magntic field $h/J=0.1,0.01,0.001$ for $L=128$.}
    \label{fig:LH_DMRG}
\end{figure}
\section{L and h dependence on the order parameter $\os$}
\label{app:c}

We systematically studied the effect of system size L and the field h near the transition point. We show the $\os$ for the two interaction strengths, U=0.0 and 1.5, and vary the frustration strength across the transition point. The weak dependence on system size indicates that the results converge in the thermodynamic limit. Panels (c) and (d) illustrate the behavior of the $\os$ with local magnetic field $h/J=$0.1, 0.01, and 0.001. The nearly overlapping curves across different field strengths indicate that the magnetic field has a negligible effect on $\os$.

\begin{figure}[tbh]  
     \includegraphics[width=1.0\columnwidth]{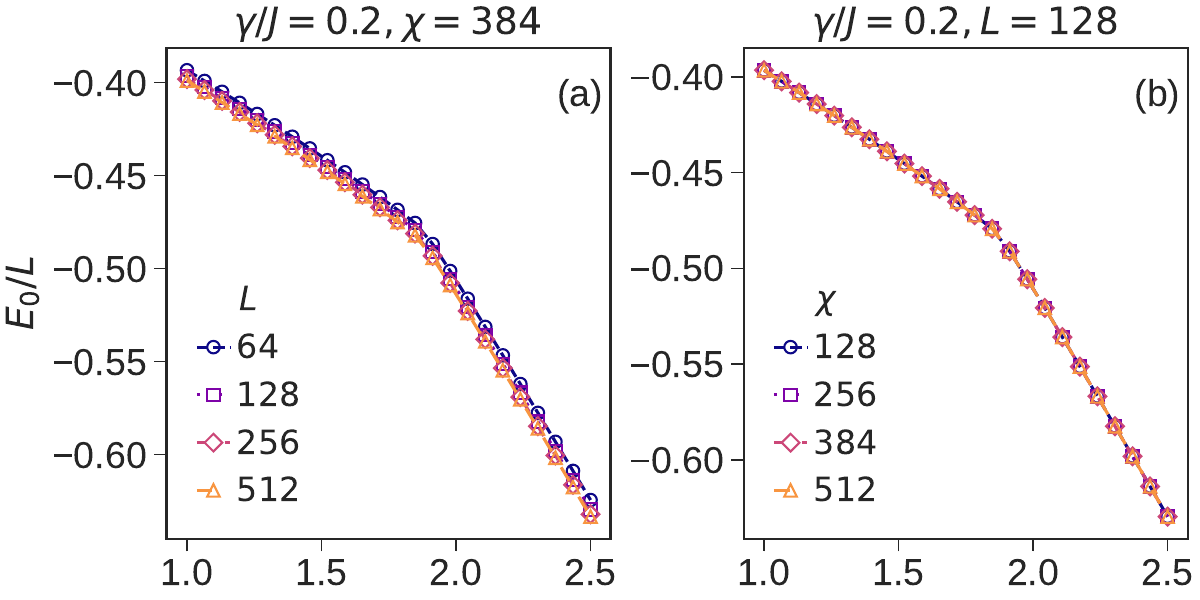}
    \caption{(a) Ground state energy normalized by system size for $L=64, 128, 256, 512$. Bond dimension dependence shows in (b) for fixed $L=128$ and $\chi=128,256,384,512$. $\gamma/J$ is kept fixed at 0.2 in both cases.}
    \label{fig:CHI_DMRG}
\end{figure}
\section{Convergence of ground state energy  with $\chi$}
\label{app:d}

Further, we investigate finite-size effects by computing the ground state energy, $E_0/L$, as a function of the interaction strength $U$ for several system sizes and bond dimensions $\chi$. Panel (a) of Fig.~\ref{fig:CHI_DMRG} shows the ground-state energy $E_0$ for $\gamma=0.2$, obtained using a fixed bond dimension $\chi=384$, for system sizes $L=64, 128, 256,$ and $512$. The almost perfect collapse of the energy curves for different $L$ indicates that finite-size effects are already negligible for the system sizes considered. Combined with the weak dependence of $E_0$ on the bond dimension $\chi$, this demonstrates that our DMRG calculations are well converged and reliably capture the thermodynamic-limit behavior of the model.

\section{L and $\chi$ dependence of the emergence of order in the absence of Ising interaction}
\label{app:e}
\begin{figure}[tbh]  
     \includegraphics[width=1.0\columnwidth]{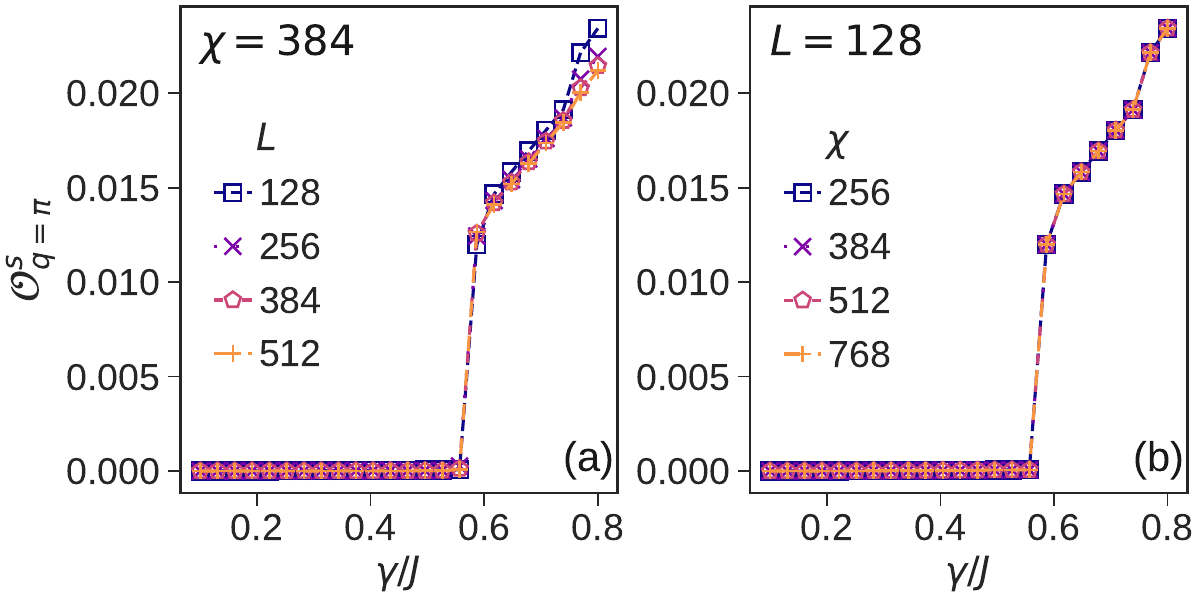}
    \caption{(a) $\os$ for$ L=128, 256, 384,512$ for fixed $\chi=384$. (b) For fixed L=128 and $\chi=256,384,512,768$. $U/J$ is kept fixed to 0.0 in both cases.}
    \label{fig:u0case}
\end{figure}

We perform a finite-size analysis of the order parameter $\os$ using DMRG simulations.  Panel (a) of Fig.~\ref{fig:u0case} shows the system-size dependence of $\os$, while panel (b) presents its bond-dimension dependence. As shown in the figure, the order parameter exhibits a sharp onset beyond a critical $\gamma/J$, and its magnitude remains nearly unchanged with increasing system size $L$. The numerical data for different $L$ collapse onto each other with only minor deviations, indicating that the observed order is robust.

\section{Ground state energy analysis in case of ferromagnetic interaction}
\label{app:f}
\begin{figure}[tbh]  
    \includegraphics[width=0.6\columnwidth]{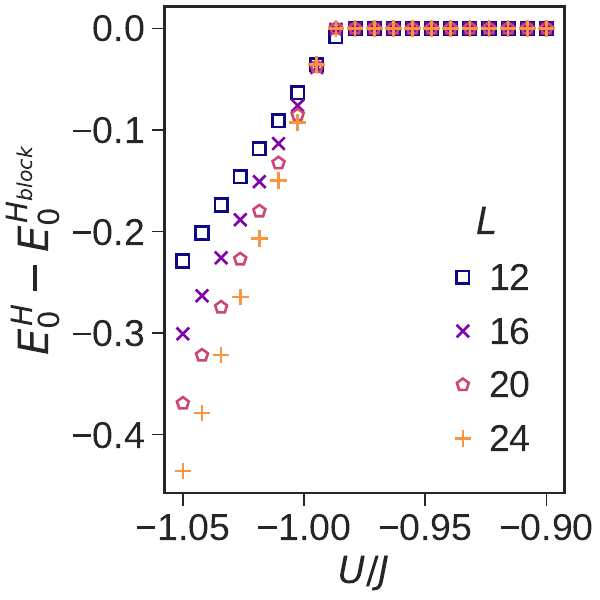}
    \caption{Difference between the ground state energies $E^{H}_{0}$ and $E^{H}_{Block}$ for L=12,16,20,24 as funtion of interaction strength with $\gamma/J=0$.}
    \label{fig:Egs1}
\end{figure}

We computed the ground state energy of the Hamiltonian given by Eq.\eqref{eq:ham} and mapped the Hamiltonian Eq.~\eqref{eq:H_block}, with ground state energy denoted by $E^{H}_{0}$ and $E^{H}_{Block}$ respectively. Fig.~\ref{fig:Egs1} shows the $E^{H}_{0}-E^{H_{Block}}_{0}$ for different system sizes with $\gamma/J=0$. The results clearly indicate that the mapping to the XXZ model remains valid only up to $U \approx -0.98$, beyond which noticeable deviations arise, indicating the breakdown of the mapping.

%apsrev4-2.bst 2019-01-14 (MD) hand-edited version of apsrev4-1.bst
%Control: key (0)
%Control: author (8) initials jnrlst
%Control: editor formatted (1) identically to author
%Control: production of article title (0) allowed
%Control: page (0) single
%Control: year (1) truncated
%Control: production of eprint (0) enabled
%

%\bibliography{bibliography.bib}
\typeout{get arXiv to do 4 passes: Label(s) may have changed. Rerun}
\end{document}